\documentclass[doublespacing]{elsart}
\usepackage[english]{babel}

\usepackage{epsfig}
\usepackage{graphics}
\usepackage{cite}

\usepackage{amssymb}

\begin{document}

\begin{frontmatter}

\title{Diffraction of waves by screens (apertures in
 screens) with time-varying dimensions. Time-varying Kirchhoff's integral
 representation for moving boundaries}

\author{V.G. Baryshevsky}

\address{Research Institute for Nuclear Problems, Belarusian State
University, 11~Bobruiskaya Str., Minsk 220030, Belarus}
\ead{bar@inp.bsu.by, v\_baryshevsky@yahoo.com}

\begin{abstract}
The diffraction of electromagnetic waves by screens (apertures in
screens) with time-varying dimensions is studied.  The generalized
vector Kirchhoff's representation for this case is obtained. It is
also shown that with accuracy up to the  terms of the order of $
\frac{v}{c} << 1 $,  the expressions for the scattered wave  and
instantaneous power can be derived from the appropriate
expressions for a stationary case by substituting the
time-dependent parameters of the screen dimensions (e.g.
time-dependent radius) for constant parameters of screen
dimensions (e.g., the screen radius)  appearing in the formulas
describing the stationary case.
\end{abstract}

\end{frontmatter}

\section{Introduction}

 Diffraction of electromagnetic waves by different bodies and
 screens is the subject of multiple studies \cite{jack,solimento}. The most important
 theoretical results  on diffraction
 by metal screens and  apertures  were reported by
 Kirchhoff, Bethe, and Bouwkamp (see, for example, \cite{jack,bethe,bouwkamp}).
 Recent research
 into  electromagnetic wave transmission
 through an array of small holes \cite{nikitin}
 has  renewed the  interest in electromagnetic wave diffraction
 \cite{nikitin,carretero,garcia}.
 It should be mentioned that all the papers cited above considered the
 diffraction of electromagnetic waves by metal screens
and  apertures  with time-constant dimensions.

 In this paper we present -- to my knowledge -- the first theoretical study of diffraction of electromagnetic waves by screens (apertures in
 screens) with time-varying dimensions and obtain the generalized vector Kirchhoff representation for this case.
Such diffraction case may occur, e.g., when the electromagnetic
wave is incident on the aperture made in a metal screen by laser
piercing, when the electromagnetic wave traverses the  cylindrical
metal
 shell collapsed under a  Z-pinch, or when
 a metal wire irradiated with the electromagnetic wave is exploded
 under the action of a power current pulse running across it.

The paper is arranged as follows: Section 2 generalizes the scalar
Kirchhoff's integral representation to the case of  screens and
apertures with time-dependent dimensions.  Section 3 discusses
wave diffraction by the screen's circular aperture with
time-dependent radius. Section 4 studies diffraction by a sphere
with time-dependent radius. Section 5 derives the time-dependent
vector Kirchhoff's representation describing the electromagnetic
wave diffraction by a screen (aperture) with time-dependent
dimensions.

\section{Generalized scalar Kirchhoff's integral representation  for the
case of  screens and apertures with time-dependent dimensions}

Let an electromagnetic wave be incident on a screen (aperture)
with time-dependent dimensions.  We shall first recall Green's
formula. According to  Gauss's flux theorem,  for any vector field
$\vec A(\vec r)$ in the volume $V$ enclosed by the surface $S$
there holds the equality \cite{morse}
\begin{equation}
\label{gauss} \int_{V} {\texttt{div}}\,\vec A(\vec r) d^3 r =
\oint_S \vec A\vec n dS,
\end{equation}
where $ \vec n$ is the unit normal vector to the surface, which is
directed outside the volume.

 Following \cite{jack}, we shall begin the consideration from the diffraction of a scalar field,
 with  $\psi$ denoting one of the electromagnetic-field
 components. (It should be mentioned that for the case of scalar wave
 propagation,
 the extension of the Kirchhoff's
scalar formula to apply to moving surfaces in acoustics was
obtained in \cite{morgans,farassat}.)  Let $\vec A =\varphi
\vec{\nabla}\psi$, where $\varphi$ and $\psi$ are the arbitrary
scalar functions.
Then
\begin{equation}
\label{sca1} \texttt{div} (\varphi \vec{\nabla}\psi)=\varphi
 \Delta \psi + \vec{\nabla}\varphi\cdot \vec{\nabla}\psi
 \end{equation}
 and
 \begin{equation}
\label{sca2} \varphi(\vec{\nabla}\psi)\cdot \vec n =
\varphi\frac{\partial\psi}{\partial n} = \varphi\vec
n\vec{\nabla}\psi,
 \end{equation}
where $\frac{\partial}{\partial n}$ is the derivative on the
surface $S$ taken along the direction of the outer normal relative
to the volume $V$.

Let us substitute (\ref{sca1}) and (\ref{sca2}) into
(\ref{gauss}). After certain transformations (for details see,
e.g., \cite{jack}), we  obtain the equality called Green's theorem
\cite{jack}:
\begin{equation}
\label{green} \int_{V} (\varphi \Delta \psi - \psi \Delta \varphi)
d^3 r = \oint_{S}\left[\varphi\frac{\partial \varphi}{\partial n}
- \psi\frac{\partial \varphi}{\partial n}\right] dS.
\end{equation}

Let us assume that the volume $V$ and the surface $S$ are
time-independent. In this case, according to \cite{jack}, we can
recast the wave equation        

\begin{equation}
\label{wave} \Delta \psi - \frac{1}{c^2}\frac{\partial^2 \psi}
{\partial t^2}= - 4 \pi f(\vec r, t)
 \end{equation}
 in the integral form ($c$ is the speed of light, $f(\vec r, t)$ is the distribution density of the
 sources) that enables us to write the solution  of the wave
 equation for $\psi( r,t)$ using explicitly  the initial conditions   $\psi(\vec
 r,t_0)$ and $\frac{\partial \psi(\vec r, t)}{\partial t}|_{t>t_0}$, as well as  the boundary conditions on the
surface \cite{jack}. This integral form of wave equation is known
as the scalar Kirchhoff's integral representation.

Because in the case under consideration the volume $V$ and the
surface $S$ are time-dependent, we need to choose a different
method of obtaining the  integral equation.

 We shall integrate
the left- and right-hand sides of (\ref{green}) between the
 time limits $[t_0,t_1]$ (compare with
\cite{jack} ). In this case, we can obtain from (\ref{green})
 \begin{equation}
 \label{e6}
 \int_{t_0}^{t_1} dt^\prime \int_{V(t^\prime)} d^3 r^\prime
 (\varphi\Delta\psi -\psi\Delta\varphi) =  \int_{t_0}^{t_1}
 dt^\prime \oint_{S(t^\prime)}\left(\varphi\frac{\partial
 \psi}{\partial n^\prime}-\psi\frac{\partial \varphi}{\partial
 n^\prime}\right) dS.
 \end{equation}

We shall further assume that $\psi =\psi$ and $\varphi = G$, where
$G$ is the Green function of the wave equation
\begin{equation}
\label{wave2}
\left(\Delta_r-\frac{1}{c^2}\frac{\partial^2}{\partial t^2}\right)
G(\vec r, t; \vec r\,^\prime, t^\prime) = - 4\pi \delta (\vec r-
\vec r\,^\prime) \delta(t-t^\prime).
\end{equation}
The upper limit of integration, $t_1$, is chosen to be greater
than $t$: $t_1>t$. As a result, we can write:
\begin{eqnarray}
\label{e8} \int_{t_0}^{t_1} dt^\prime \int_{V(t^\prime)} d^3
r^\prime
\left[ G(\vec r, t; \vec r\,^\prime, t^\prime)\Delta_{r'}
\psi(\vec r\,^\prime, t^\prime) - \psi(\vec r\,^\prime, t^\prime)
\Delta_{r'}^\prime G(\vec r, t; \vec r\,^\prime, t^\prime) \right]
= \nonumber\\
\int_{t_0}^{t_1} dt^\prime \int_{S(t^\prime)} \left[
G(\vec r, t; \vec r\,^\prime, t^\prime)\frac{\partial \psi(\vec
r\,^\prime, t^\prime)}{\partial n^\prime} -\psi(\vec r\,^\prime,
t^\prime) \frac{\partial}{\partial n^\prime} G(\vec r, t; \vec
r\,^\prime, t^\prime) \right]
d S^\prime.
\end{eqnarray}
Let us recall that in the right-hand side of equation (\ref{e8})
all points $\vec r\,^\prime$ lie on the the surface $S(t^\prime)$
enclosing the volume $V(t^\prime)$; $\frac{\partial \psi}{\partial
n^\prime}=\vec n\,^\prime \frac{\partial\psi}{\partial \vec r
\,^\prime}$.
We shall make use of (\ref{wave}) and (\ref{wave2})
and recast (\ref{e8}) as follows:
\begin{eqnarray}
\label{e9} & &\int_{t_0}^{t_1} dt'\int _{V(t')}d^3 r'\left\{G(\vec
r, t;\vec r\,', t')\left[\frac{1}{c^2}\frac{\partial^2\psi(\vec
r\,',
t')}{\partial t'^2}- 4\pi f(\vec r\,', t')\right]\right.\nonumber\\
& & - \left.\psi(\vec r\,',
t')\left[\frac{1}{c^2}\frac{\partial^2}{\partial t'^2}G(\vec r, t;
\vec r\,', t')- 4\pi \delta(\vec r - \vec r\,')\delta(t-t')\right]\right\}\\
& & =  \int_{t_0}^{t_1} dt' \int _{S(t')} \left[G(\vec r, t; \vec
r\,',
 t')\frac{\partial\psi(\vec r\,', t')}{\partial n'} - \psi(\vec r\,',
 t')\frac{\partial}{\partial n'}G(\vec r,  t; \vec r\,', t')\right]
 dS', \nonumber
\end{eqnarray}
that is,
\begin{eqnarray}
\label{e10} & & 4\pi \psi(\vec r, t)  -  4\pi \int_{t_0}^{t_1} dt'
\int_{V(t')} d^3 r' G(\vec r, t; \vec r\,', t') f (\vec r\,', t') + \\
& + & \int^{t_1}_{t_0} dt' \int _{V(t')} d^3 r' \left\{G(\vec r,
t; \vec r\,', t')\frac{1}{c^2}\frac{\partial^2 \psi(\vec r\,',
t')}{\partial t'^2}- \psi(\vec r\,',
t')\frac{1}{c^2}\frac{\partial^2}{\partial t'^2}G(\vec r, t; \vec r\,', t')\right\} = \nonumber\\
& = & \int_{t_0}^{t_1} dt' \int _{S(t')} \left[G(\vec r, t; \vec
r\,',
 t')\frac{\partial\psi(\vec r\,', t')}{\partial n'} - \psi(\vec r\,',
 t')\frac{\partial}{\partial n'}G(\vec r,  t; \vec r\,', t')\right]
 dS'\nonumber
\end{eqnarray}
By changing the order of integration over $d t'$ and $d^3 r'$, $d
t'$ and $d S'$ we can obtain Kirchhoff's integral representation
(see \cite{bouwkamp,nikitin}), i.e., the integral equations for $
\psi$, where the function $\psi$ is expressed in terms of the
values of $\psi $ and its coordinate and time derivatives on the
surface $S$.
But this change  of the  order of integration is
valid only in the case when the volume $V$ and the surface $S$ are
time-independent! If $V=V(t)$ and $S=S(t)$, the order of
integration cannot be changed.
 For this reason, we shall
introduce the function $\Theta(\vec r\,', \vec R(t'))$ such that
$\Theta(\vec r\,', \vec R (t'))=1$ for all points $\vec r'$
contained in the volume $V(t')$ enclosed by the surface $S(t')$
that is described by the coordinates $\vec R (t')$ of the points
belonging to the surface; $\Theta(\vec r\,', \vec R (t'))=0$ for
points $\vec r\,'$ exterior to the volume $V(t')$.
Using this function, we can write the integral $\int_{_{V(t')}}
d^3 r' ...$ over the time-dependent volume $V(t')$ in the third
term on the left-hand side of (\ref{e10}) in the form of the
integral $\int d^3 r' \Theta(\vec r\,', \vec R(t'))...$, where
integration is performed over a certain large (infinite in the
limit) constant volume.
As a result, we can perform time integration by parts on the
left-hand side of (\ref{e10}) and transform the second-order
derivatives into the first-order ones: $\frac{\partial^2}{\partial
t'^2} \longrightarrow \frac{\partial}{\partial t'}$.
This gives us the integral equation for the function $\psi(\vec r,
t)$ expressed in terms of the function $f(\vec r, t)$, describing
the radiation source, $\psi$ and the derivatives thereof on the
surface enclosing the volume $V(t)$, as well as in terms of the
initial value of $\psi(\vec r, t_0)$ and the derivative thereof at
the initial time. The equation thus obtained generalizes
Kirchhoff's integral representation to a nonstationary case.

\section{Diffraction of waves  by the screen's circular aperture with
time-dependent radius a(t)}

We shall start our further consideration with the case of wave
diffraction by the screen's circular aperture with time-dependent
radius $a(t)$. Let a perfectly conducting planar screen of
thickness $L$ be placed in the plane $x,y$; the $z$-axis being
orthogonal to the plane of the screen (Fig. 1).

\begin{figure}[htp]
\label{fig1}
\begin{center}
       \includegraphics[height=6.5cm]{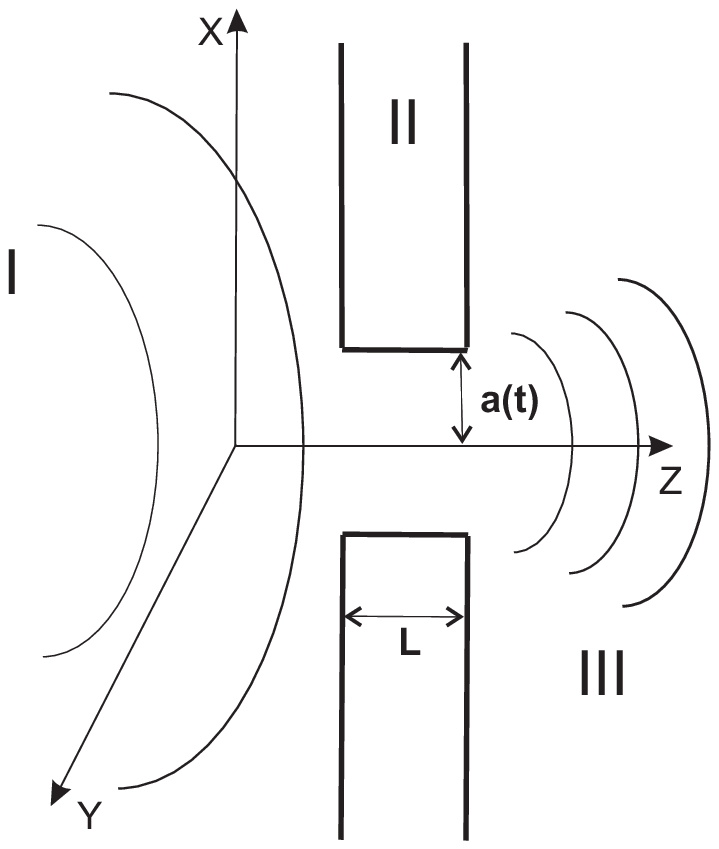}
\caption{}
\end{center}
\end{figure}

The radiation sources are placed on the left of the screen, i.e.,
in the region $z<0$. We are concerned with the field produced by
the source in space in the presence of the screen.

As we mentioned in the Introduction, a detailed  analysis for a
thin screen with a time-independent aperture radius was given in
\cite{bethe,bouwkamp} and generalized to the case with screen of
thickness $L$ in \cite{nikitin}. Here we shall consider the case
when the screen thickness is much less than the incident radiation
wavelength. In this case, we can rule out that part of the volume
which is occupied by region II with varying radius.
Let us separately consider  regions I and II that have an
invariable volume. For these regions, the volume $V(t)$ and the
surface $S(t)$ in (\ref{e9}) and (\ref{e10})  are
time-independent.  As a result, we can perform time integration
over $t'$ by parts in the right-hand side of the equation, and the
integral equation for $\psi(\vec r, t)$ takes a well-known form
\cite{jack}:

\begin{eqnarray}
\label{e11} & & \psi(\vec r,t)  =  \int_{t_0}^{t_1} d t' \int d^3
r'
G(\vec r, t; \vec r\,', t') f(\vec r\,',t')+ \\
& + & \frac{1}{4\pi c^2} \int d^3 r' \left.\left.\left (G(\vec r,
t; \vec r\,', t_0)\frac{\partial \psi (\vec r\,', t')}{\partial
t'}\right|_{t'=t_0} - \psi(\vec r\,', t_0)\frac{\partial}{\partial
t'} G(\vec r, t;
\vec r\,', t')\right|_{t'=t_0}\right)+\nonumber\\
& + & \frac{1}{4\pi} \int_{t_0}^{t_1} d t' \int_S \left[G(\vec r,
t; \vec r\,', t')\frac{\partial\psi (\vec r\,', t')}{\partial n'}
- \psi (\vec r\,', t')\frac{\partial}{\partial n'} G(\vec r, t;
\vec r\,', t')\right] dS'\nonumber.
\end{eqnarray}
For further consideration it will be useful to recall that Green's
function $G(\vec r, t; \vec r\,', t')$  of wave equation
(\ref{wave}) satisfies  (\ref{wave2}) and has the form
\begin{equation}
\label{e12} G(\vec r, t; \vec r\,', t')=\frac{\delta
\left(t-t'-\frac{|\vec r- \vec r\,'|}{c}\right)}{|\vec r- \vec
r\,'|}.
\end{equation}
The presence of $\delta$ function in Green's function enables us
to perform time integration in (\ref{e11}), too. According to
(\ref{e11}), the value of $\psi (\vec r, t)$ at any point in the
volume $V$ is defined by the function $f(\vec r, t)$ of the
source, the value of $\psi(r,t_0)$, and
$\left.\frac{\partial\psi(r,t)}{\partial t}\right|_{t=t_0}$ at the
initial time, as well as by the value of $\psi(r,t)$ on the
surface $S$ \cite{jack}.

Let us consider (\ref{e11}) in region  III on the right of the
screen. No radiation source is placed in this region, and so there
is  no field present at the initial time.

Using the explicit
expression for Green's function, we can derive the following
equation for $\psi(\vec r,t)$ (for details, see \cite{jack}):
\begin{eqnarray}
\label{e13} & & \psi(\vec r, t)  = \\
& =&  \frac{1}{4\pi} \oint_{S}\vec n \left[\frac{1}{R}\vec \nabla
_{\vec r\,'}\psi (\vec r\,', t') -\frac{\vec r - \vec r\,'}{|\vec
r-\vec r\,'|^3} \psi (\vec r\,', t')-\frac{\vec r-\vec
r\,'}{c|\vec r-\vec r\,'|^2}\frac{\partial\psi(\vec r\,',
t')}{\partial t'}\right]_{{del}} d S', \nonumber
\end{eqnarray}
where $\left[...\right]_{del}$ indicates that after  the
derivatives are taken, $t'$ is assumed to be the delay time
$t'=t-\frac{|\vec r- \vec r\,'|}{c}$.  Thus to find $\psi (\vec r,
t)$, we need to know  $\psi$ and its derivative, which certainly,
cannot be taken arbitrary \cite{jack}, but requires the solution
of the problem with given initial and boundary conditions.

Let a wave packet $\psi_0(\vec r, t)$ be incident on the screen
from the left (see Fig.):
\begin{equation}
\label{e14} \psi_0(\vec r, t)=\int A(\vec k - \vec k_0)e^{i\vec k
\vec r} e ^{-i\omega(k) t} d^3 k.
\end{equation}
When the characteristic wavelength $\lambda$ of the packet is much
less than the aperture radius $a(t)$,  following Kirchhoff, we
shall assume  that in the same way as in the stationary case,
$\psi(\vec r\,', t')$ and $\frac{\partial\psi(\vec
r\,',t')}{\partial t'}$ equal zero on the surface of the screen in
the region $z=0 $ of screen location, while in the aperture
$\psi(\vec r\,', t')=\psi_0(\vec r\,', t')$ and
$\frac{\partial\psi(\vec r\,',t')}{\partial
t'}=\frac{\partial\psi_0(\vec r\,',t')}{\partial t'}$. (To avoid
confusion, hereafter the velocity $v$ of the aperture radius
change is assumed to be $v\ll c $; below is shown how to consider
the relativistic effects in diffraction of electromagnetic waves.)

As a result, at a large distance from the screen, $R\gg a$,
equation (\ref{e13}) for the region on the right of the screen
takes the form
\begin{eqnarray}
& & \psi(\vec r, t)= \\
& = & \frac{1}{4\pi} \int_0^{a({t'_a})}\,\int^{2\pi}_0 \rho'
d\rho' d\varphi \, \vec n \left[\frac{1}{R}\vec \nabla_{\vec
r\,'}\psi_0(\vec r\,', t') -\frac{\vec R}{R^3}\psi_0(\vec r\,',
t') -\frac{\vec R}{cR^2}\frac{\partial\psi_0(\vec r\,',
t')}{\partial t'}\right], \nonumber \label{e15}
\end{eqnarray}
where $t'_a=t-\frac{|\vec r -a (t_a')|}{c}$.

Let us recall here that $R=|\vec r-\vec r\,'|$ and
$t'=t-\frac{|\vec r - \vec r\,'|}{c}$.
 We shall further consider  $\psi (\vec r, t)$ at a distance
$r\gg a$ and  discard the second term $\sim\frac{1}{R^2}$, leaving
only the terms proportional to $\frac{1}{R}$.

Let us substitute (\ref{e14}) for $\psi_0(\vec r,t )$ into
 (\ref{e15}). Now we have
\begin{eqnarray}
\label{e16} \psi(\vec r, t)=\int d^3 k A(\vec
k-\vec{k}_0)\frac{1}{4\pi r} \int_0^{a({t'_a})}\,\int^{2\pi}_0
\rho' d\rho' d\varphi \, \vec n \left[i\vec{k} +
i\frac{\omega(k)}{c}\vec
n_r\right]e^{i\vec k\vec r'} e^{-i\omega(k) t'}\nonumber\\
\\
= -\frac{ik_0}{4\pi r}(1+\cos\vartheta) \int d^3\kappa A(\vec
\kappa) \int_0^{a({t'_a})}\,\int^{2\pi}_0 \rho' d\rho' d\varphi
\left\{ e^{i\vec{\kappa}\vec{\rho}\,'} e^{+i\vec k_0\vec{\rho}\,'}
e^{-i\omega(\vec k_0+\vec{{\kappa}}) t'}\right\},\nonumber
\end{eqnarray}
where $\vec{n}_r=\frac{\vec{r}}{r}$, $\vartheta$ is the angle
between the $z$-axis and the direction of $\vec r$, and
$t'=t-\frac{|\vec r-\vec r\,'|}{c}$.

Let us consider the diffraction of a quasi-monochromatic wave
packet whose transverse dimensions are much greater than the
aperture diameter and whose amplitude is, hence, almost constant
in the aperture region. In this case {${\kappa_{0\perp} a\ll 1}$}.
Then we have
\begin{eqnarray}
\label{e17} & & \psi(\vec r, t)= -\frac{i k_0}{4\pi r} A\left(
t-\frac{r}{c}\right)(1+\cos
\vartheta)\int_0^{a({t'_a})}\,\int_0^{2\pi} d^2\rho' e^{-i(\vec
k_0'-\vec k_0)\vec\rho\,'} e^{i k_0 r}e^{-i\omega_0 t}\nonumber\\
& & = \left(-\frac{i k_0}{4\pi}\right)(1+\cos
\vartheta)\int_0^{a(t-\frac{r}{c})}\,\int_0^{2\pi} d^2\rho'
e^{-i(\vec k_{\perp}'-\vec k_{0\perp})\vec\rho\,'} \frac{e^{ik_0
r}}{r}
e^{-i\omega_0 t} A\left(t-\frac{r}{c}\right)\nonumber\\
& & = - \frac{i k_0}{4\pi}
(1+\cos\vartheta)\int_0^{a(t-\frac{r}{c})} \rho' d\rho' d\varphi
e^{-i q\rho\cos\varphi} A\left(t-\frac{r}{c}\right)
\frac{e^{ik_0r}}{r} e^{-i\omega_0 t},
\end{eqnarray}
\begin{equation}
\label{e18} \vec q = \vec k_{0\perp}'- \vec k_{0\perp}, \,\,
q=|\vec q|.
\end{equation}
We have
\begin{eqnarray}
\label{e19} \psi(\vec r, t)=-\frac{i
k_0}{2}(1+\cos\vartheta)\int_0^{a(t-\frac{r}{c})}\rho d\rho J_0
(q\rho) A\left(t-\frac{r}{c}\right)\frac{e^{ik_0 r}}{r} e^{-i\omega_0 t}\\
=-\frac{ik_0}{2}(1+\cos \vartheta)
a\left(t-\frac{r}{c}\right)\frac{J_1[a{\left(t-\frac{r}{c}\right)}q]}{q}
\frac{e^{i k_0 r}}{r} e^{-i\omega_0
t}A\left(t-\frac{r}{c}\right).\nonumber
\end{eqnarray}
Let us recall that $t>\frac{r}{c}$.

Expression (\ref{e19}) for the  wave diffracted by the aperture
with a time-dependent radius includes the term $ a\left(
t-\frac{r}{c}\right)$ instead of $a$ appearing in  the expression
for $\psi (r,t)$, which describes wave diffraction by the aperture
with invariable $a$. The same occurs in the case of diffraction by
the additional screen: in formulas describing diffraction by the
additional screen, the time-independent size is replaced by a
time-dependent one. As  a result, in the scalar theory the
instantaneous output power radiated by the aperture per unit solid
angle can be written in a form much similar to that for the
stationary case (compare \cite{jack},\cite{nikitin})
\begin{equation}
\label{e20} \frac{d P}{d\Omega}=const\left[k_0
a\left(t-\frac{r}{c}\right)\right]^2\left|\frac{J_1\left(a\left(t-\frac{r}{c}\right)q\right)}{a\left(t-\frac{r}{c}\right)q}\right|^2
\left|A\left(t-\frac{r}{c}\right)\right|^2,
\end{equation}
\[
q=|\vec k_{0\perp}'-\vec k_{0\perp}|^2.
\]
According to (\ref{e19}), the signal that has passed through the
aperture appears to be modulated. Let us pay attention to the fact
that for large $a q$, the Bessel functions are proportional to the
sum of $\cos$ and $\sin$, i.e., they are proportional to $e^{\pm
ia\left(t-\frac{r}{c}\right)q}$. If the radius $a (
t-\frac{r}{c})$ increases at a constant rate $v_a$, then it
follows from (\ref{e19}) that the function $\psi(\vec r, t)$
appears the be proportional to $e^{i(r_0\pm\frac{v_a}{c}q)r}
e^{-i(\omega_0\pm v_a q)t}$, and a shift in the  frequency $
\Delta\omega=\pm v_q q$ arises.

\newpage

\section{Diffraction by a sphere of radius $a(t)$}

 As another example we shall consider the diffraction of
waves by a sphere of radius $a(t)$ undergoing radial contraction
or expansion (Fig. 2).
\begin{figure}[h!]
\label{fig2}
\begin{center}
     \resizebox{60mm}{!}
       {\includegraphics{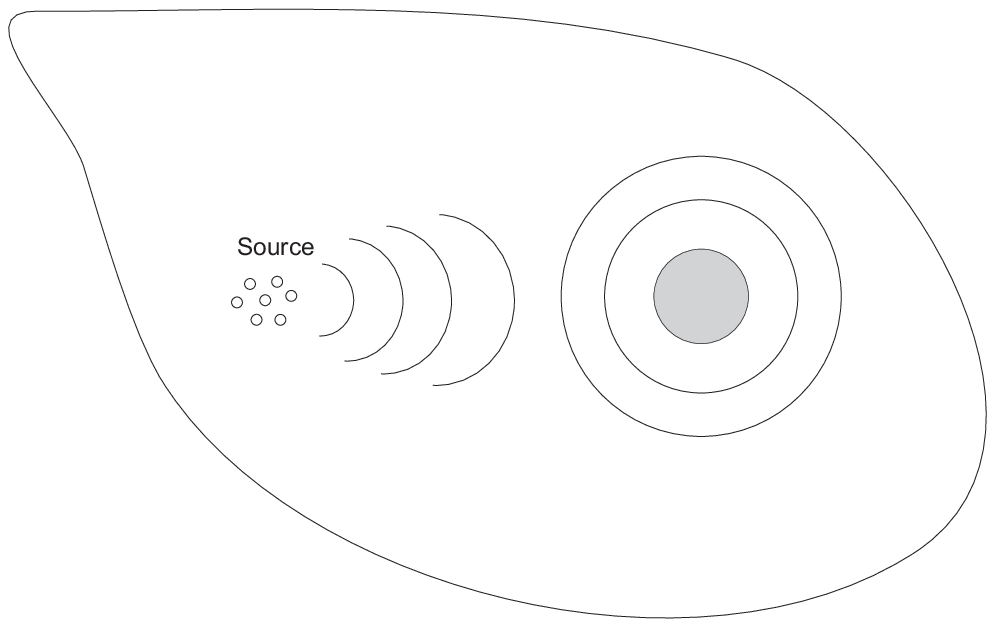}}\\
\caption{}
\end{center}
\end{figure}

 According to the rules stated earlier in
Section 2, in this case we can write (\ref{e10}) in the form:
\begin{eqnarray}
\label{in1}
& & \psi(\vec r, t)  - \int_{t_0}^{t_1} dt'\int_{V_{sorce}} d^3 r'
G
(\vec r,t; \vec r\, ' t')f(\vec r\,',t') + \\
& + & \frac{1}{4\pi c^2}\int d\Omega'\int_0^{\infty} r'^2 d
r'\int_{t_0}^{t_1}dt' \times \nonumber \\
& \times & \left\{\Theta(r'-a(t'))G(\vec r,t; \vec r\, ' t')
\frac{\partial^2\psi(\vec r\,',t')}{\partial t'^2}-
\Theta(r'-a(t'))\psi(\vec r\,',t')\frac{\partial^2 G(\vec r,t;
\vec r\, ' t') t}{\partial
t^2}\right\} \nonumber\\
& = & \frac{1}{4\pi}\int_{t_0}^{t_1} dt' \int _{S(t')}
\left[G(\vec r, t; \vec r\,',
 t')\frac{\partial\psi(\vec r\,', t')}{\partial n'} - \psi(\vec r\,',
 t')\frac{\partial}{\partial n'}G(\vec r,  t; \vec r\,', t')\right]
 dS'.\nonumber ,
\end{eqnarray}
where $d\Omega'$ denotes integration over the solid angle of
vector $r\,'$.

After partial integration of the third component on the left-hand
side of (\ref{e11}), we have
\begin{eqnarray}
\label{in5} & & \psi(\vec r, t) =
\int^{t_1}_{t_0}\int_{V_{source}}dt'd^3r'G(\vec r,t; \vec r\, '
t')
f(\vec r\,',t')-  \\
& - & \frac{1}{4\pi c^2}\int^{t_1}_{t_0} d t'\int
d\Omega'a^2(t')v(t')\left(G\frac{\partial\psi}{\partial
t'}-\psi\frac{\partial G}{\partial t'}\right)\nonumber\\
& + & \frac{1}{4\pi} \int ^{t_1}_{t_0} d t'\int d\Omega'
a^2(t')\left\{G(\vec r, t; \vec r\,' , t')(\vec n
\vec\nabla_{r'})\psi(\vec r\,',t')-\psi(\vec r\,',t')(\vec n
\vec\nabla_{r'})G(\vec r, t; \vec r\,' , t')\right\}, \nonumber
\end{eqnarray}

where $v(t')=\frac{da(t')}{dt'}$ is the velocity of sphere
expansion (contraction).

Let us make use of the fact that \cite{bouwkamp}
\[
G(\vec r, t; \vec r\,'
t')=\frac{\delta\left(t-t'-\frac{R}{c}\right)}{R}; \]
\[
\vec\nabla_{r'} G=\frac{(\vec r-\vec
r\,')}{|r-r'|^3}\delta\left(t'+\frac{R}{c}-t\right)-\frac{(\vec
r-\vec r\,')}{cR^2}\delta'\left(t'+\frac{R}{c}-t\right); \]

\noindent $\vec R= \vec r- \vec r\,'$; here $\delta'$  is the
derivative of the $\delta$-function.

As a result, we have
\begin{eqnarray}
\label{in9} \psi(\vec r, t) & = & \psi_{source}+\frac{1}{4\pi}
\int d\Omega' a^2(t')\left\{\frac{1}{|\vec r -\vec r (t')|}\vec
n\vec\nabla_{r'} \psi(\vec r\,', t')\right.\\
& - & \left.\psi(\vec r\,', t')\frac{\vec n (\vec r- \vec
r\,')}{|\vec r-\vec r\,'|^3}-\frac{\vec n(\vec r-\vec r\,'
(t'))}{c|\vec r-\vec r\,'(t')|^2}\frac{\partial\psi(\vec r,
t')}{\partial
t'}\right\}\nonumber\\
& - & \frac{1}{4\pi}\int d\Omega\frac{\partial}{\partial
t'}\left\{a^2(t')\frac{\vec n(\vec r- \vec r \,' (t'))}{c|\vec
r-\vec r\,' (t')|^2}\right\}\psi(\vec r\,',t')\nonumber\\
& + & \frac{1}{2\pi c^2}\int d\Omega' a^2 (t')v(t')\frac{1}{|\vec
r -\vec
r\,'(t')|}\frac{\partial\psi}{\partial t'}\nonumber\\
& + & \frac{1}{4\pi c^2}\int d\Omega' \frac{1}{|\vec r -\vec
r\,'(t')|}\frac{\partial}{\partial t'}(a^2(t')V(t')) \psi(\vec
r\,',t').\nonumber
\end{eqnarray}

At a large distance from the sphere (\ref{in9}) simplifies as
follows
\begin{eqnarray}
\label{in10} & & \psi(\vec r,t) =  \psi _{source}+\frac{1}{4\pi
r}\int d \Omega ' a^2 (t')\left\{\vec n\vec \nabla_{r'}\psi(\vec
r\,',t')-\frac{\vec n \vec n_{r}}{c}\frac{\partial\psi (\vec
r\,',t')}{\partial t'}\right\} - \nonumber\\
& - & \frac{\vec n \vec n_r}{2\pi c r}\int d\Omega'a(t') v(t')
\psi(\vec r\,',t') + \frac{1}{2\pi c^2 r}\int d\Omega
a^2(t')v(t')\frac{\partial\psi(\vec r\,',t')}{\partial t'}
+ \nonumber \\
& + & \frac{1}{4\pi c^2 r}\int d\Omega \frac{\partial}{\partial
t'}(a^2(t'))v(t'))\psi(\vec r\,', t').
\end{eqnarray}
Thus, according to (\ref{in9}) and (\ref{in10}), the time-varying
dimensions of the sphere lead to the appearance of additional
terms (third, fourth, and fifth), whose amplitudes depend on the
velocity $v$ and  the acceleration of sphere expansion
(contraction). The values of these additional contributions are
proportional to the ratio of the velocity $v$ to the speed of
light $c$. Because the power of the scattered wave $\psi_{sc}$ is
proportional to $|\psi_{sc}|^2$, these additional contributions
can be observed in studying the interference of the contributions
to the intensity of scattered radiation that come from the product
of the second and third (fourth or fifth) terms. At the same time,
the contribution coming to the intensity from the additional terms
themselves at $\frac{v}{c}\ll 1$ is small  and can be omitted.

\section{Vector analogue of the time-dependent Kirchhoff's integral
representation}

Let us now proceed to the electromagnetic wave diffraction by a
screen (aperture) with time-dependent dimensions. At first glance
it may seem that because the electromagnetic field has a vector
character, in this case only the coefficients appearing in
(\ref{in1})--(\ref{in10}) will change. But the situation appears
to be more complicated, as  due to relativistic effects, the
magnetic field also makes a contribution to the strength of the
electric field responsible for the current running in the moving
screen. As a result, the contribution coming  from the magnetic
field to the process of electromagnetic field penetration through
the screen can play an important, even critical role, for example,
when an apertured  screen is placed in the near-induction zone of
the magnetic dipole.

To consider diffraction of the electromagnetic field in the case
of time-dependent dimensions, we shall make use of the vector
analogue of Green's theorem \cite{morse}. Following \cite{morse},
for vector analogue of Green's theorem, we can write the
relationship between the two vectors $\vec P$ and $\vec F$ in the
form:
\begin{eqnarray}
\label{e21} & & \int_V \left\{\vec P\cdot \vec\nabla^2\vec F -
\vec F
\cdot\vec\nabla^2\vec P\right\} d^3 v\\
 & =& \oint\left\{(\vec P \,{\texttt{div}} \,\vec F-\vec F\,
{\texttt{div}}\, \vec P) \cdot \vec n-(\vec P\cdot[\vec n\times\,
{\texttt{rot}}\vec F] +\,{\texttt{rot}}\,\vec P\cdot[\vec n\times
\vec F])\right\} ds, \nonumber
\end{eqnarray}
where $\vec n$ is the  external normal unit vector to  the surface
and

$\vec \nabla^2 \vec F = \texttt{grad}\,\texttt{div} \,\vec F -
\texttt{rot}\,\texttt{rot}\,\vec F$.

We shall further assume that vector $\vec P$ is the vector to be
found, i.e., either the electric field strength $\vec E$ or the
magnetic field strength $\vec H$. The Green function, which in the
considered case is a tensor, is used for vector $\vec F$. The
propagation of an electromagnetic wave in a free space is
described by the vector Helmholtz equation \cite{morse}
\begin{equation}
\label{e22} \vec \nabla^2 \vec P
-\frac{1}{c^2}\frac{\partial^2\vec P}{\partial t^2}= - 4\pi \vec
Q_P(\vec r, t).
\end{equation}
The Green function is a symmetrical tensor $(\hat{G}\cdot \vec P=
\vec P \cdot \hat{G})$  and has the form \cite{morse}
\begin{equation}
\label{e23} \vec \nabla^2\hat{G} (\vec r, t; \vec r\,' t')
-\frac{1}{c^2}\frac{\partial ^2 \hat{G}}{\partial t^2}= -4\pi
\hat{I}\delta(\vec r -\vec r\, ')\delta(t-t'),
\end{equation}
where $\hat{I}$ is the unit operator.

Using (\ref{e22}) and (\ref{e23}), we can recast (\ref{e21}) as
follows:
\begin{eqnarray}
\label{e24} & & \int_{V(t)}\left\{\vec P{(\vec r\,',
t')}\cdot\left(\frac{1}{c^2}\frac{\partial^2 \hat{G}(\vec r, t;
\vec r\,' t')}{\partial t^2}- 4 \pi\hat{I} \delta (\vec r - \vec r
') \delta (t - t')\right) - \right. \\
& -& \left. G(\vec r, t; \vec r\,'
t')\cdot\left(\frac{1}{c^2}\frac{\partial^2 \vec P{(\vec r\,',
t')}}{\partial t^2}- 4
\pi \vec Q(\vec r ', t')\right)\right\} d^3 V =\nonumber\\
&=& \oint_{S(t')}\left\{(\vec P \,\texttt{div}\, \hat{G}-\hat{G}\,
\texttt{div}\, \vec P)\cdot \vec n - (\vec P \cdot[\vec n\times\,
{\texttt{rot}}\hat{G}] +\,{\texttt{rot}}\,\vec P\cdot[\vec n\times
\hat{G}])\right\} ds,\nonumber
\end{eqnarray}
Let us integrate (\ref{e24}) over the time $t'$ between the limits
from $t_0$ to $t_1>t$. As a result, we have  the following
integral equation for the vector field $\vec P(\vec r, t)$ in the
case of time-dependent $V(t)$ and $S(t)$ (let us recall that the
retarded Green function $\hat{G}$ equals zero for $t'>t$):
\begin{eqnarray}
\label{e25} & & \vec P (\vec r, t)  =
\int^{t_1}_{t_0}\int_{V_{(t')}} \hat{G}(\vec r, t; \vec r\,'
t')\vec Q_P (\vec r\,', t') d^3 v dt' \\
& - & \frac{1}{4\pi
c^2}\int^{t_1}_{t_0}\int_{V(t')}\left(\hat{G}(\vec r, t; \vec r\,'
t')\frac{\partial^2 \vec P (\vec r \,' ,t')}{\partial t'^2}-\vec
P(\vec r\, ', t') \frac{\partial^2 \hat{G}(\vec r, t;
\vec r\,' t')}{\partial t'^2}\right) d^3 r' d t'\nonumber\\
& - & \frac{1}{4\pi}\int_{t_0}^{t_1}\oint_{S_(t')}\left\{(\vec P
\,\texttt{div}\, \hat{G}-\hat{G}\, \texttt{div}\, \vec P)\cdot
\vec n - (\vec P \cdot[\vec n\times\, {\texttt{rot}}\hat{G}]
+\,{\texttt{rot}}\,\vec P\cdot[\vec n\times \hat{G}])\right\}
ds,\nonumber
\end{eqnarray}

Integral relation (\ref{e25}) generalizes the vector Kirchhoff's
integral representation with time-independent $V$ and $S$ to the
case with time-dependent volume $V(t)$ and  surface $S(t)$.

The integral relation (\ref{e25}) obtained here, as well as in the
scalar case, can be used to find the fields $\vec E$ and $\vec B$
for the EM wave diffraction by an object (screen) with moving
boundaries when the boundary conditions on the surface $S$ are
fulfilled.
%
In contrast to a static case, in the discussed case of moving
boundaries account should be taken of the fact that relativistic
effects mix the fields $\vec E$ and $\vec B$ on the surface $S$,
making them  dependent on the speed of the boundary motion.

The secondary waves appearing through diffraction  are generated
by those charges moving in the body which are set in motion by the
Lorentz force that depends on both electric and magnetic
components of  the incident electromagnetic wave.
In a weakly relativistic case, the currents $\vec J$ excited on
the surface of the  conducting body are proportional to $\vec E
+\frac{1}{c}[\vec v(t) \vec B]$. As a result, for example, on the
moving boundary of a high-conductivity metal, the tangential
component of the effective electric field $(\vec E
+\frac{1}{c}[\vec v(t) \vec B])_t$ equals zero rather than that of
the electric field.

It is noteworthy that the vector analogue of  time-dependent
Kirchhoff's integral representation, derived here, can be obtained
from a scalar representation if by the field $\psi$ we understand
the Cartesian components of the electric or magnetic field and
then perform vector addition of the derived equations.
However, thus obtained equations  are inconvenient
 for further use, since the boundary conditions on the
surface $S$ are difficult to satisfy, and therefore need
modifying. To do this,  we shall use the same approach as in
deriving the vector analogue of  Kirchhoff's integral
representation in the case of time-independent $V$ and $S$,  given
in \cite{jack}.

For ease of treatment  we further drop the additional terms
appearing in (\ref{e10}) due to the varying dimensions of the
screen that are proportional to $ \frac{v}{c}$.

According to (\ref{e10}), in the scalar case the field $\psi(\vec
r, t)$ in the volume $V$ (in the absence of the sources inside $V$
and zero initial values of $\psi$ and $\frac{\partial
\psi}{\partial t})  $ is described by the expression of the form
\begin{eqnarray}
\label{e27} \psi(\vec r,t)  & = &  \frac{1}{4\pi}\int ^{t1}_{t_0}
dt'\oint_{S(t')} \left\{ G(\vec r, \vec r\, ', t, \vec t')(\vec n
\vec
\nabla\,')(\psi(\vec r\,', t')\right.\\
 & - & \left. \psi(\vec r\,', t')(\vec n \vec\nabla') G (\vec r,
\vec r\,', t, t')\right\} d S'. \nonumber
\end{eqnarray}

If by $\psi$ we understand a certain Cartesian component of the
electric $\vec E$ or magnetic $\vec B$ field, then we can write
\begin{equation}
\label{e28} \vec E(\vec r, t) =\frac{1}{4\pi}\int_{t_0}^{t_1} d t'
\oint_{S(t')} \left\{G(\vec n \vec \nabla ') \vec E - \vec E (\vec
n \vec \nabla ') G \right\} dS'.
\end{equation}
The integral over the surface converts  to the form \cite{jack}
\begin{eqnarray}
\label{e29} & & \oint_S \left\{G(\vec n \vec \nabla ') \vec E   -
\vec E
(\vec n \vec \nabla ') G \right\} dS' \\
& =&   -\oint_S\left\{(\vec n \vec E)\vec \nabla ' G + \left[[\vec
n\times \vec E]\times \vec \nabla ' G \right] + G [\vec n\times
 rot'\vec E]\right\} dS'. \nonumber
\end{eqnarray}
As a result, we have the following relationship:
\begin{eqnarray}
\label{e30} \vec E (\vec r, t) & = & -
\frac{1}{4\pi}\int_{t_0}^{t_1} d t' \oint_{S(t')}\left\{\left(\vec
n\vec E(\vec r\,', t')\right)\vec\nabla 'G +\left [[\vec n\times
\vec E]\times
\vec\nabla \,' G\right]\right.\\
& -& \left. G\frac{1}{c}\left[\vec n\times \frac{\partial \vec B
(\vec r\,' t')}{\partial t'}\right]\right\} dS.\nonumber
\end{eqnarray}
In a similar manner, we  obtain for $\vec B$
\begin{eqnarray}
\label{e31} \vec B (\vec r, t) & = & -
\frac{1}{4\pi}\int_{t_0}^{t_1} d t' \oint_{S(t')}\left\{\left(\vec
n\vec B(\vec r\,', t')\right)\vec\nabla 'G +\left [[\vec n\times
\vec B]\times
\vec\nabla \,' G\right]\right.\\
& + & \left. G\frac{1}{c}\left[\vec n\times \frac{\partial \vec E
(\vec r\,' t')}{\partial t'}\right]\right\} dS'.\nonumber
\end{eqnarray}

Equations (\ref{e30}) and (\ref{e31}) derived here are a specific
case of more generalized expressions  (\ref{e25}), derived earlier
in this section. Like (\ref{e25}), they are valid in the case when
the surface $S$ is time-dependent, i.e., when $S=S(t')$. For
further transformations, let us make use of the fact that
\begin{eqnarray}
\label{e32} \vec \nabla\,' G(\vec r, \vec r\, ' ; t, t') & = &-
\frac{\vec R}{r}\frac{\partial}{\partial R}\left[\frac{\delta
(t'+\frac{R}{c}-t)}{R}\right] \\
& =&-\frac{\vec
R}{R}\left\{-\frac{\delta(t'+\frac{R}{c}-t)}{R^2}+\frac{\delta'(t'+\frac{R}{c}-t)}{cR}\right\}\nonumber\\
& = & \frac{\vec R}{R^3}\delta(t'+\frac{R}{c}-t) -\frac{\vec
R}{cR^2}\delta'(t'+\frac{R}{c}-t).\nonumber
\end{eqnarray}
Here $\vec R= \vec r - \vec r\,'$.

As a result, we have
\begin{eqnarray}
\label{e33} \vec E (\vec r, t) &=& -\frac{1}{4\pi}\int_{t_0}^{t_1}
dt'\oint_{S(t')}\left\{(\vec n \vec E (\vec r\,', t
'))\left(\frac{\vec R}{R^3}\delta(t'+\frac{R}{c}-t)-\frac{\vec
R}{cR^2}\delta'(t'+\frac{R}{c}-t)\right)\right.\nonumber\\
& + & \left.\left[[\vec n\times\vec E]\times \left(\frac{\vec
R}{R^3}\delta(t'+\frac{R}{c}-t)-\frac{\vec
R}{cR^2}\delta'(t'+\frac{R}{c}-t)\right)\right]\right.\nonumber\\
& - & \left.\frac{1}{Rc}\delta(t'+\frac{R}{c}-t)\left[\vec n
\times \frac{\partial\vec B (\vec r\,', t)}{\partial
t'}\right]\right\} ds.
\end{eqnarray}
Upon integration of (\ref{e33}) over time using $\delta$-function,
we  drop the terms proportional to $\frac{v}{c}$ and obtain the
following equation
 \begin{eqnarray}
 \label{e34}
 & & \vec E(\vec r,t) =- \frac{1}{4\pi}\oint_{S(t')}\left\{\frac{\vec
 R}{R^3}(\vec n\vec E(\vec r\,',t'))+\frac{\vec R}{c
 R^2}\frac{\partial}{\partial t'}(\vec n\vec E(\vec r\,', t'))\right.\\
 & + & \left. \left[[\vec n\times \vec E] \times \frac{\vec
 R}{R^3}\right]+\frac{1}{c}\left[\left[\vec n\times\frac{\partial
 E}{\partial t'}\right]\times \frac{\vec R}{R^2}\right]-\frac{1}{R
 c}\left[\vec n\times \frac{\partial B(\vec r\,', t')}{\partial
 t'}\right]\right\} ds.\nonumber
 \end{eqnarray}
Here $t'=t-\frac{|\vec r- \vec r\,'|}{c}$.

As $r\rightarrow \infty$, we can discard the terms proportional to
$\frac{1}{R^2}$ and the integral over that part of the surface
which is located a long distance away from the screen. In this
case, we have
\begin{eqnarray}
\label{e35} \vec E(\vec r, t) & = & - \frac{1}{4\pi c
r}\\
& \times & \oint_{S(t')}\left\{\frac{\vec r}{r}\left(\vec n \frac
{\partial \vec E (\vec r\, 't')}{\partial t'}\right) +
\left[\left[\vec n \times \frac{\partial\vec E}{\partial
t'}\right]\times\frac{\vec r}{r}\right]  - \left[\vec n \times
\frac{\partial \vec B}{\partial t'}\right]\right\} d s.\nonumber
\end{eqnarray}
Here $\vec n$ is the outer normal. If the normal is directed
towards the observation area, the sign of the expression should be
reversed. Then we have
\begin{eqnarray}
\label{e36} \vec E & = &   \frac{1}{4\pi c
r}\oint_{S(t')}\left\{\frac{\vec r}{r}\left(\vec n \frac{\partial
\vec E}{\partial t'}\right) + \left[\left[\vec n \times
\frac{\partial \vec E}{\partial t}\right]\times\frac{\vec
r}{r}\right]- \left[\vec n\times\frac{\partial\vec B}{\partial
t'}\right]\right\}ds,\\
t'& = & t-\frac{R}{c}.\nonumber
\end{eqnarray}
A similar to (\ref{e36}) expression for the magnetic field $\vec
B(\vec r, t)$ follows from (\ref{e31}):
\begin{equation}
\label{e37} \vec B = \frac{1}{4\pi c
r}\oint_{S(t')}\left\{\frac{\vec r}{r}\left(\vec
n\frac{\partial\vec B}{\partial t'}\right) +\left[\left[\vec n
\times \frac{\partial \vec B}{\partial t'}\right]\times\frac{\vec
r}{r}\right] +\left[\vec n\times \frac{\partial \vec E (\vec
r\,',t')}{\partial t'}\right]\right\} ds.
\end{equation}
For monochromatic fields $\vec E$, $\vec B$  $\sim e^{-i\omega t}$
and $t$-independent $S$,  expression (\ref{e33}) converts to a
well-known stationary expression (see formula (9.115) in
\cite{jack}):
\begin{equation}
\label{e37v2} \vec E(\vec r)=\frac{e ^{ikr}}{4\pi i
r}\oint_S\left\{\vec k(\vec n\vec E(\vec r\,'))+\left[[\vec
n\times \vec E]\times \vec k\right]-k[\vec n\times \vec B(\vec
r\,')]\right\}e^{-i\vec k\vec r\,'} ds.\\
\end{equation}
Let us recall that $\vec k=k\frac{\vec r}{r}$.

In a similar manner, we have for a magnetic field
\begin{equation}
\label{e38} \vec B =\frac{e^{ikr}}{4\pi i r}\oint\left\{\vec
k(\vec n \vec B(\vec r\,'))+\left[[\vec n\times \vec B]\times \vec
k\right]+ k [\vec n\times \vec E(\vec r\,',
\omega)]\right\}e^{-i\vec k\vec r\,'} ds,
\end{equation}
i.e.,
\begin{equation}
\label{e39} \vec B(\vec r)= \frac{e^{ik\vec r}}{4\pi i r}\vec
k\times \int_S\left[[[\vec n\times \vec E]\times\vec n_r]-[\vec
n\times \vec B]\right]e^{-i\vec k\vec r\,'} d s.
\end{equation}
We shall also recall that
\[
\vec B=\frac{1}{k}[\vec k\times \vec E].
\]

Let a quasi-monochromatic wave packet be scattered by a perfectly
conducting screen with dimensions much greater than the
wavelength; the wave packet time length $T_p\gg \frac{L}{c}$,
where $L$ is the characteristic dimension of the screen. In a
similar manner as was done in the scalar case, we shall write the
incident wave packet in the form:
\begin{equation}
\label{e42} \vec E_{in}(\vec r, t)=\vec E_0\int A(\vec k-\vec k_0)
e^{i\vec k\vec r} e^{-i\omega t} d^3 k.
\end{equation}

Let us recall, that in view of the above, the motion of the screen
boundary results in the appearance of the additional terms
proportional to $\frac{v}{c}$ and $\frac{dv}{dt}$. These terms not
only change the amplitudes of the fields but also lead to the
electric field contribution to the magnetic field (and vice versa)
due to relativistic effects.

In the beginning, we shall neglect these contributions. Let us
recall that {${\omega_0\gg\frac{1}{T}}$}, where $T$ is the
characteristic time during which the speed of the screen changes.
Using (\ref{e42}) and the boundary conditions on the screen
surface \cite{jack}, at a large distance from the screen we can
obtain the following expression for the scattered fields $\vec
E_S(\vec r,t)$ and $\vec B_S (\vec r,t)$:
\begin{eqnarray}
\label{e43} \vec E_S (\vec r,t) & = & \frac{e^{ikr} e^{-i\omega_0
t}}{4\pi i r} A(r-ct) \times \\
& \times & \oint_{s(t')}\left\{\vec k(\vec n\vec E_{0S}) +
\left[[\vec n\times \vec E_{0S}]\times \vec k\right]-k[\vec
n\times \vec B_{0S}]\right\}e^{-i(\vec k-\vec k_0)\vec r\,'}
ds,\nonumber
\end{eqnarray}
where  $\vec B_S=[\vec n_k\times \vec E_S]$, $\vec k=k_0\frac{\vec
r}{r}$, and the fields (see\cite{jack}) $ \vec E_{0S}\approx -\vec
E _0$ and $\vec B_{0S}\approx -\vec B_0$ are in the shadow region
of the obstacle. In the illuminated region of the obstacle, we
have
\begin{eqnarray}
\label{e444} & & \vec n \vec E_{0S}\simeq\vec n \vec E_0, \,\,\vec
n\times \vec B_{0S}=\vec n\times \vec B_0,\,\, \vec n\times \vec
E_{0S}=-\vec n
\times \vec E_0, \\
& & \vec n\vec B_{0S}=-\vec n\vec B_0,\,\,\vec B=\frac{1}{k}[\vec
k\times\vec E]. \nonumber
\end{eqnarray}
Using (\ref{e37v2}), we can write a similar expression for the
magnetic field $\vec B(\vec r, t)$
\begin{eqnarray}
& & \vec B_S(\vec r,t)=-\frac{i}{4\pi r} e^{i
k_0r}e^{-i\omega_0 t} A(r-ct) \times \\
& \times &
 \oint_{S(t')}\left\{\vec k(\vec n \vec
B_{0S})+\left[[\vec n\times \vec B_{0S}]\times\vec k\right]+
k_0[\vec n\times\vec E_{0S}]\right\} e^{-i(\vec k-\vec k_0)\vec
r\, '} ds, \label{e44} \nonumber
\end{eqnarray}

i.e.,
\begin{equation}
\vec B(\vec r,t)=\vec F_B \frac{e^{i k_0r}}{4\pi i r} e^{-i w_0t}
A(r-ct)
\end{equation}
where

\begin{equation}
\vec F_B=\vec k\times \vec f= e^{i k_0r}\vec k\times
\oint_{S(t')}\left\{\left[[\vec n\times \vec E_{0S}]\times \vec
n_k\right]-[\vec n\times\vec B_{0S}]\right\}ds.
\end{equation}

In a similar manner as in the scalar case, (\ref{e43})  and
(\ref{e45}) can be obtained by substituting $S(t')$ for $S$ in the
expression describing scattering of a monochromatic wave by a
screen with time-independent dimensions and then multiplying the
expression by the amplitude $A(r-ct)$ of the wave packet.

As a result, in a similar manner as in the stationary case, from
(\ref{e43}) we obtain, for example, the following expression  for
the electric field strength in the wave scattered at a small
angle:
\begin{equation}
\label{e45} \vec E_S\approx
ika^2(t')\frac{I_1(ka(t')\vartheta)}{ka(t')\vartheta}\frac{e^{i(kr-\omega_0
t)}}{r}\frac{(\vec k\times\vec E_0)\times \vec k}{k^2}A(r-ct),
\end{equation}
where $\vartheta$ is the scattering angle.

As we  noted earlier, this contribution to  field scattering can
be obtained from the expression for the field in a stationary case
\cite{jack}-- \cite{garcia} by substituting $a(t')$ for $a$ and
multiplying the expression for the field by the amplitude
$A(r-ct)$ of the wave packet.

Let us recall that in deriving  (\ref{e34})--(\ref{e36}), we
discarded the terms proportional to $ \frac{v}{c}$ and to
acceleration $ \frac{dv}{dt}$.  Similar to the scalar case, the
contribution from these terms can be observed experimentally by
studying the interference pattern of scattered radiation.

It is noteworthy that the additional terms can be divided into
three groups. The first group includes the terms proportional to $
\frac{v}{c}$ that only change the moduli of the fields. The second
group comprises the terms proportional to the acceleration $
\frac{dv}{dt}$, and the third one contains the terms admixing the
magnetic field to the electric (or vice versa). Despite the
smallness of $ \frac{v}{c}$, the third group of terms is
fundamentally important  when the electric field itself is induced
by the motion of the conductor in the magnetic field (i.e.,
electromagnetic induction).

In this case, the expressions for $\vec{E_0}$  and  $\vec{B_0}$ in
(\ref{e444})  should be rewritten for a relativistic case. For
example, in a weakly relativistic case $\vec{E_0}$ converts to
$\vec{E_0} + \frac{1}{c} [\vec{v}\vec{B_0}]$, and $\vec{B_0}$ to
$\vec{B_0} - \frac{1}{c} [\vec{v}\vec{E_0}]$. If the field
$\vec{E_0}$ is small,  the rewritten expression for the initial
fields includes $\vec{E_0}\simeq \frac{1}{c}[\vec{v}\vec{B_0}] $
and $\vec{B_0}=\vec{B_0} $.

Here we will not write  cumbersome expressions describing the
contribution to the  intensity and the angular distribution of
radiation  that  comes from the additional terms, leaving their
detailed consideration for specific  cases of practical
importance.

\section{Conclusion}
We generalized Kirchhoff's vector  representation to the case of
screens (apertures) with time-dependent dimensions. It has been
shown that in the case when $\frac{v}{c}\ll 1$, the expressions
for the scattered wave and instantaneous power can be derived from
the appropriate expressions for a stationary case
\cite{jack,solimento,bethe,bouwkamp,nikitin,carretero} by
substituting the time-dependent screen dimensions (e.g.
time-dependent radius) for constant screen dimensions (e.g., the
screen radius) appearing in the formulas describing the stationary
case.


\begin{thebibliography}{10}

\bibitem{jack} J. Jackson,   \emph{Classical Electrodynamics} (3rd
ed.) New York: John Wiley  \& Sons, 1998.

\bibitem{solimento} S. Solimeno, B. Crosignani, P. DiPorto, \emph{Guiding, Diffraction and Confinement of Optical
Radiation}  New York, Academic, 1986.

\bibitem{bethe} H. A. Bethe, Theory of Diffraction by Small Holes, \emph{Phys. Rev. Lett.}, Vol. 66 (1944) 153.

\bibitem{bouwkamp} C. J. Bouwkamp, Diffraction Theory, \emph{Rep. Prog. Phys.}, Vol. 17 (1954)  35.

\bibitem{nikitin} A.~Yu.~Nikitin, D.~Zueco, F.~J.~Garcia-Vidal and
L.~Martin-Moreno, Electromagnetic wave transmission through a
small hole in a perfect electric conductor of finite thickness,
\emph{Phys. Rev. B } Vol. 78 (2008) 165429.

\bibitem{carretero} S. Carretero-Palacios, F. J. Garcia-Vidal, L.
Martin-Moreno, Sergio G. Rodrigo, Effect of film thickness and
dielectric environment on optical transmission through
subwavelength holes, \emph{Phys. Rev. B } Vol. 85 (2012)  035417.

\bibitem{garcia} F.~J.~Garcia-Vidal, Esteban~Moreno, J.~A.~Porto,
 L.~Martin-Moreno, Transmission of light through a single rectangular hole, \emph{Phys.Rev. Lett.} Vol. 95 (2005)  103901.

\bibitem{morgans} W.R. Morgans, The Kirchhoff formula extended to a moving
surface, \emph{Philosophical Magazine}, Vol. 9 (1930)  141.


\bibitem{farassat} F. Farassat, M. K. Myers, Extension of
Kirchhoff's formula to radiation from moving surfaces,
\emph{Journal of Sound and Vibration}, Vol. 123 (3) (1988) 451.
\bibitem{morse} Ph.~Morse and H.~Feshbach, \emph{Methods of Theoretical Physics}, Part 1, McGraw-Hill, 1953.




\end{thebibliography}
\end{document}